\documentstyle[psfig]{mn}

\input amssym.def
\input amssym

\newcommand{\ds}{\displaystyle}
\newcommand{\be}{\begin{equation}}
\newcommand{\en}{\end{equation}}
\begin{document}
\title{A critical-density closed Universe in Brans-Dicke Theory}

\author[Sergio del Campo and Norman Cruz]
{Sergio del Campo$^{1,3}$ and Norman Cruz$^{2,4}$ \\
$^1$Instituto de F\'{\i}sica, Universidad Cat\'{o}lica de Valpara\'{\i}so, Av. Brasil 2950,
Valpara\'{\i}so, Chile.\\
$^2$Dept. de F\'{\i}sica, Universidad de Santiago de Chile, Casilla 307,
Correo 2, Santiago, Chile.\\
$^3$ sdelcamp@ucv.cl\\
$^4$ ncruz@lauca.usach.cl\\}\maketitle


\begin{abstract}

In a Brans-Dicke (BD) cosmological model, the energy
density associated with some scalar field decreases as
$\displaystyle a^{{-2}\left( \frac{\omega _{o}+ {\frac12}%
}{\omega _{o}+1}\right)} $ with the scale factor $a(t)$ of the
Universe, giving a matter with an Equation of state $\displaystyle
p=-\frac{1}{3}\left( \frac{2+\omega _{o}}{1+\omega _{o}}\right)
\rho $. In this model, the Universe could be closed but still have
a nonrelativistic-matter density corresponding to its critical
value, $\Omega _{o}=1$. Different cosmological expressions, such
as, luminosity distance, angular diameter, number count and ratio
of the redshift tickness-angular size, are determined in terms of
the redshift for this model.
\end{abstract}



\section{Introduction}

In the Friedman-Robertson-Walker (FRW) model, there are enough
evidences that the Universe is flat, with a large component of
negative pressure. This component was consider at first to be
just the cosmological constant (or vacuum energy)
(Tuner et al. 1984; Peebles 1984; Ostriker \& Steinhard 1995;
Liddle et al. 1996). Another
possibility was to consider topological deffects (Vilenkin 1984),
but, the one which has receiving a great deal
of attention today, is related to a scalar field, $Q$,
the so-called ``quintessence'' model or the QCDM model
(Caldwell, Dave \& Steinhardt 1998).
This scalar field is characterized by a very
negative pressure, i.e., $ P_{Q}=w\rho _{Q}$, with $w\leq -1/3$.

Measurements of distant SNe Ia (at $ z\sim 1$) indicate that the
expansion of the Universe is in accelaration rather than
deceleration (Perlmutter et al. 1998).

According to Garnavich et al. (1998) this is consistent with the
existence of an unknown component for the energy density, which
could be considered to correspond to the quintessence component.
On the other hand, a test of the standard model, including
spacetime geometry, galaxy peculiar velocities, structure
formation, and early Universe physics, supports, in many of these
cases, a flat Universe model with the presence of a cosmological
constant (Peebles 1998). Specially, a model in which a short
period of inflation occurs at a very early epoch in the evolution
of the Universe.Most of these models predict that the total
density parameter, $\Omega_o$, be unity.

Given the idea that the Universe could be described by a flat
geometry, an interesting question to ask is whether this flatness
could be due to a local effect. This sort of question has been
considered in the literature (\cite{KaTo}). There, a $k=1$ model
was taken into account, together with a total density parameter
corresponding to $\Omega_{o}<1$. In this model, openness is
obtained by considering a matter component whose equation of state
is $\displaystyle p=-\frac{1}{3}\rho $. This sort of state of
Equation has been reported in models in which topological defects,
such as texture or tangled strings, are important components of
the total energy of the Universe.

In this paper we analyse a FRW closed model, $k=1$, filled with
dust and a effective density energy component characterized by a
negative pressure, using the Brans-Dicke (BD) theory (\cite{BrDi})
with a potential associated to the BD field. In particular, we
investigate the conditions in order to have a model with positive
curvature which mimics a flat universe at low redshift. This imply
to determine the contributions of the scalar field $Q(t)$ and the
BD field to the total energy density, which cancel the
contribution due to curvature $\Omega_{ok}=-1/a_{0}H_{0}$. We
should note that the obtained model is the generalization of the
Einstein-de Sitter model ($k=0$) in the context of the BD theory.
Our model is far from being realistic, since it gives an age for
the universe,$t_{0}$, very close to $2H_{0}/3$, and a deceleration
parameter, $q_{0}$, very close to $1/2$. These values, as we
mentioned above, desagree with the measurements of distant
supernovae. Nevertheless, we clearly pointed out that our proposal
is to set the conditions under a closed universe mimics a flat
universe in the BD theory. At the end of the following section we
shall briefly discuss how in the BD theory a flat universe with
negative pressures, presents acceleration in agreement with actual
observations.

We studied this sort of model in a previous paper (\cite{CrdCHe}),
but that analysis was limited to the model's intrinsic
characteristics, such that the explicit determination of the
scalar fields, $Q(t)$, its potential $V(Q)$ and the BD potential \
$V(\Phi )$, where $\Phi$ represents the BD scalar field. In this
paper we investigate the model's cosmological characteristics,
such that, the proper distance to the horizon, the luminosity
distance, the angular size, the differential number of galaxies,
and the ratio $(1/z)\delta z/\delta \theta $, where, all of these
quantities are given as a function of the redshift $z$. We compare
these parameters with that corresponding to the Einstein-de Sitter
model.

We should note that our model, in the limit $\omega_{0}
\rightarrow \infty$ and $\Phi = const$, gives rise to the $k=0$
Einstein-de Sitter model.


\section{ Characteristics of the model}

Assuming homogeneity and isotropy, the FRW metric for a closed
universe is
\be
\ds d\,{s}^{2}\,=\,\,d\,{t}^{2}\,-\,
\,a(\,{t}\,)^{2}\,d\Omega_{k=1}^2,
\en
with $d\Omega_{k=1}^2$ representing the spatial line element
asociated to the hypersurfaces of homogeneity, corresponding to a
three sphere. $a(t)$ represents the  scale factor, which together
with the assumption that the $Q$ scalar field is homogeneous,i.e.,
$Q=Q({t})$, we obtain the fundamental field Equations of the BD
model, given by (where the dots representing derivatives with
respect to time $t$. We use units in which $ c=\hbar =1 $ ) $$
  \hspace{-3.5cm}\ds H^{2}+H\left( \frac{\stackrel{\cdot }{\Phi
}}{\Phi }\right) = \frac{\omega _{o}}{6}\left(
\frac{\stackrel{\cdot }{\Phi }}{\Phi }\right) ^{2}+ $$
\be
\label{H1} \hspace{3.5cm} \ds \frac{8\pi }{3\Phi }\left( \rho
_{M}+ \rho _{Q}\right) -\frac{1}{a^{2}}+\frac{U\left( \Phi \right)
}{6\Phi },
\en
$$
\hspace{-3.5cm}\stackrel{\cdot \cdot }{\Phi }+3H\stackrel{\cdot }{\Phi }+
\frac{\Phi ^{3}}{
2\omega _{o}+3}\frac{d}{d\Phi }\left( \frac{U(\Phi )}{\Phi ^{2}}\right) =
$$
\be
\hspace{3.5cm} \frac{8\pi }{2\omega _{o}+3}\left[ \rho _{M}+
\left( 1-3w\right) \rho _{Q} \right], \label{Phi1}
\en
and
\begin{equation}
\stackrel{\cdot \cdot }{Q}+3H\stackrel{\cdot }{Q}=-\frac{\partial V(Q)}{
\partial Q},
\label{Q}
\end{equation}

The condition describing a model mimicing a flat Universe is
(Cruz, del Campo \& Herrera 1998) given by $$ \hspace{-3.5cm} \ds
\frac{d}{d\Phi }\left( \frac{U\left( \Phi \right) }{\Phi ^{2}}
\right) -\frac{1}{2\left( \omega _{o}+1\right) }\frac{U\left( \Phi
\right) }{ \Phi ^{3}} $$
\be
\label{co1} \hspace{3.5cm}\ds + \frac{3}{\Phi_0^2(1+\omega
_{o})}\left( \frac{\Phi_0}{\Phi }\right) ^{2\left( 2+\omega
_{o}\right) }=0,
\en
where $\Phi_0 $ is the the actual value of the BD scalar field.
Under these conditions Equation (\ref{H1}) and Equation
(\ref{Phi1})reduce to
\begin{equation}
H^{2}+H\left( \stackrel{\cdot }{\Phi }/\Phi \right) =\frac{\omega _{o}}{6}%
\left( \stackrel{\cdot }{\Phi }/\Phi \right) ^{2}+\frac{8\pi
}{3\Phi }\rho _{M}, \label{H22}
\end{equation}
and
\begin{equation}
\stackrel{\cdot \cdot }{\Phi }+3H\stackrel{\cdot }{\Phi }=\frac{8\pi }{%
2\omega _{o}+3}\rho _{M}, \label{Phi22}
\end{equation}
respectivelly. Equation (\ref{Q}) remains unaltered. Note that
this set of equations mimics a flat ($\Omega_m = \Omega_0 = 1$)
Universe.

Assuming that the matter content, $\rho_{M}$, is dominated by
dust, i.e. $\rho _{M}\sim a^{-3}$, the solutions of Equations
(\ref{H22}) and  (\ref{Phi22}) are the well known power law
solutions of a flat BD model, given by
\begin{equation}
\displaystyle a=a_o \left( \frac{t}{t_{o}}
\right)^{\frac{2}{3}\left( \frac{\omega _{o}+1}{\omega
_{o}+4/3}\right) },
\label{a}
\end{equation}
and
\begin{equation}
\ds \Phi(t)=\Phi_o \left(\frac{t}{t_o} \right)^{\left( \frac{2}
{3 \omega_o+4} \right)},
\end{equation}
where $t_{o}$ is the current age of the Universe. Using the usual formula for
the redshift $z$, $1+z=a_o/a$, we find that
\begin{equation}
\displaystyle t=\frac{2}{3}\left( \frac{\omega _{o}+1}{\omega _{o}+4/3}%
\right) H_{o}^{-1}\left( 1+z\right) ^{-\frac{2}{3}\left( \frac{\omega
_{o}+4/3}{\omega _{o}+1}\right) },  \label{t}
\end{equation}
where $H_{o}\equiv \sqrt{8\pi \rho _{M}^{o}/3\Phi _{o}} $, is the
current Hubble constant taken as $ H_{o}=100h km s^{-1} Mpc^{-1}$,
with $h$ in the range $0.6\leq h\leq 0.8$ (Bureau, Mould \&
Staveley-Smith 1996). From Equation (\ref{t}) we can obtain the
age of the Universe, $\ds t_{o}=\frac{2}{3}\left( \frac{ \omega
_{o}+1}{\omega _{o}+4/3}\right) H_{o}^{-1} = \left(
\frac{\omega_o+1}{\omega_o+4/3}\right) t_o^E$, where $t_o^E$ is
the Einstein-de Sitter value for the age of the Universe, given by
$\frac{2}{3} H_o^{-1}$. At first glance, the factor appearing in
$t_o$ seems to decrease the age of the Universe, due to $\ds
\left( \frac{\omega_o+1}{\omega_o+4/3}\right) \leq 1$. But, since
$\omega_o \geq 500$ (Reasenberg et al. 1979), this factor will be
small, and almost equal to one, and, therefore, we could use $t_o
\simeq t_o^E$.

The growth of the Q-field is given by
\begin{equation}
\displaystyle\left( \frac{\stackrel{\cdot }{Q}}{Q}\right) =2H_{o}\left(
\frac{\omega _{o}+3}{\omega _{o}+1}\right) \left( 1+z\right) ,
\end{equation}
and its present value is then determined by the values of the Hubble
constant, $H_{o}$, and the BD-parameter $\omega _{o}$. At large values of $%
\omega _{o} $ $ (\omega _{o}\longrightarrow \infty )$ it becomes $\left(
\stackrel{\cdot }{Q}/Q\right) _{o}=2H_{o}$. On the other hand, the potential
$V(Q)$ associated with this field is given by
\begin{equation}
\displaystyle V(Q)=M^{4}\left( \frac{M}{Q}\right) ^{4\left( \frac{\omega
_{o}+ {\frac12} }{\omega _{o}+1}\right) }
\end{equation}
where $M$ is a free parameter.

Since $\displaystyle\rho _{Q}=\frac{3}{2}\left( \frac{\omega
_{o}+1}{\omega _{o}+5/4}\right) V(Q)$ and the actual density
parameter $\Omega _{Q}^{\omega _{o}}$ is defined by
$\displaystyle\Omega _{Q}^{\omega _{o}}=\frac{8\pi \rho
_{Q}^{o}}{3\Phi _{o}H_{o}^{2}}$, we find that $\displaystyle\Omega
_{Q}^{\omega _{o}}=\left( \frac{\omega _{o}+1}{\omega
_{o}+5/4}\right) \frac{4\pi M^{4}}{ \Phi _{o}H_{o}^{2}}=\left(
\frac{\omega _{o}+1}{\omega _{o}+5/4}\right) \Omega _{Q}$, where
$\Omega _{Q}$ corresponds to the scalar field density parameter in
Einstein-de Sitter's model of the universe.

If we consider the same range for $\Omega_Q$ in our model, we find
that this imposes a constrain  on $M$. Since today it is required
that $ V(Q\approx M_{p\ell })\approx \rho _{m}\sim 10^{-47}
{GeV^{4}}$, where $M_{p\ell}$ is the Planck mass, we obtain
\be
\ds
M\sim \left( \rho _{m}M_{P\ell }^{4\left( \frac{\omega _{o}+ {\frac12}
}{\omega _{o}+1}\right) }\right) ^{\frac{\omega _{o}+1}{8\left( \omega _{o}+
{\frac34} \right) }}.
\en
This gives $M\sim 10^{4} {GeV}$ for $\omega
_{o}\sim 500$. This value is comparable to particle physics
scales. We mention that the evolution of $Q$ and the values of
$\Omega _{Q}$ today are very insensitive to the initial value of
$Q$, due to its attractor solution (Zlater, Wang \& Steinhardt 1998).

One point that needs to be taken into account in this kind of
model is the fraction of the Universe in causal contact. To do
so, we employ a comoving observer at coordinate $(r_{o}=0,\theta
,\varphi )$ at time $t$. A light signal satisfies the geodesic
equation of motion $ds^{2}=0$.   Therefore,   a light
signal emitted from $\left( r_{1},\theta ,\varphi \right) $ at
time $t=0$, following the line $\theta =\phi = const.$, will
reach the observer at time $t$%
\begin{equation}
\int\limits_{o}^{t}\frac{dt^{\prime }}{a(t^{\prime })}=\int
\limits_{o}^{r_{H}}\frac{dr}{\left( 1-r^{2}\right) ^{{\frac12} }},
\end{equation}
and because the proper distance to the horizon is
\begin{equation}
d_{H}(t)=\int\limits_{o}^{r_{H}}(g_{rr})^{\frac12} dr,
\end{equation}
it is found that
\begin{equation}
d_{H}(z)=a_0\,\frac{\alpha \left( \omega _{o},H_{o}\right)
}{(1+z)}\left[ 1-\left( 1+z\right) ^{-\beta \left( \omega
_{o}\right) }\right], \label{dh}
\end{equation}
where $\displaystyle\alpha \left( \omega _{o},H_{o}\right)
=\frac{2}{a_o H_{o}}\sqrt{\left( \omega _{o}+3/2\right) \left(
\omega _{o}+4/3\right) }/\left(\omega _{o}+2\right) $ and
$\displaystyle\beta \left( \omega _{o}\right) = \frac{1}{2}\left(
\frac{\omega _{o}+2}{\omega _{o}+1}\right) $. In Fig. 1 we have
plotted $d_{H}(z)$ as an function of $z$. In Einstein-de Sitter
model we have nearly obtained the same curve.


\begin{figure}
\centerline{\psfig{figure=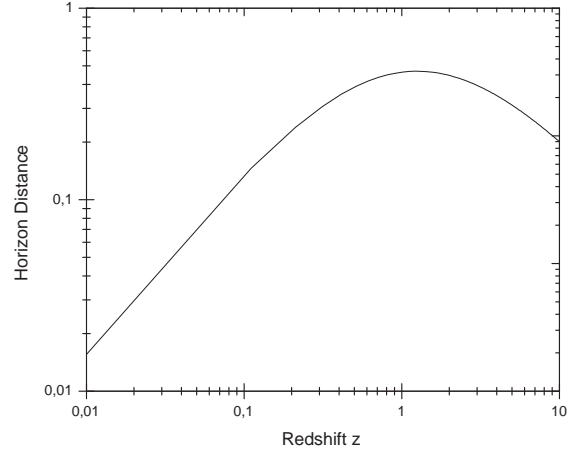,height=10.5cm} }
\caption[fig1.ps] { This plot shows the horizon distance
$d_{H}(z)$ (in unit of $ H_{o}$) as a  function of the redshift
$z$, for the BD parameter $\omega _{o}=500$. The curve coincides
with the corresponding to the Einstein-de Sitter model.
\label{fig1}}
\end{figure}


Since, $\omega _{o}\gg 1$, we can compare $d_{H}(z)$ in the BD
theory with the corresponding expression in Einstein's theory of
general relativity (Einstein-de Sitter model). Expanding Equation
(\ref{dh}) up to the first order in $1/\omega _{o}$, yields
\begin{equation}
d_{H}(z)\simeq d_{H}^{E}(z)+g(z,H_{o})(1/\omega _{o})+O(1/\omega
_{o})^{2},
\end{equation}
where $g(z,H_{o})$ becomes defined by
\begin{equation}
g(z,H_{o})=\frac{H_{o}^{-1}}{(1+z)}\left[ \frac{7}{3}\left( \frac{1}{\sqrt{%
1+z}}-1\right) +\frac{\ell n\left( 1+z\right)
}{\sqrt{1+z}}\right],
\end{equation}
and $d_{H}^{E}(z)$ represents the horizon distance in the
Einstein-de Sitter model. This expression presents a maximum for
$z\simeq 2$, and there, the difference, $\Delta d_{H}(z)\equiv
d_{H}(z)-d_{H}^{E}(z)$, computed up to order $1/\omega _{o}$,
becomes a maximum. Its value is not significant, since it is less
than one percent.

In the following we determine the deceleration  parameter $q_{o}$
for our model. This parameter is defined by $ q_{o}=-\left(
\ddot{a}/aH^{2}\right) _{o}$.  By using the solution   given by
Eq. (\ref{a}), we obtain that
\begin{equation}
q_{o}=\frac{1}{2}\left( \frac{\omega _{o}+2}{\omega
_{o}+1}\right).
\end{equation}
Note that we can write $q_{o}=\gamma \left( \omega _{o}\right)
-1$, where $ \gamma \left( \omega _{o}\right) $ is the inverse of
the exponent in the expression of the scale factor $a(t)$, i.e.
$\displaystyle\gamma \left( \omega _{o}\right) =\frac{3}{2}\left(
\frac{\omega _{o}+4/3}{\omega _{o}+1} \right) $. If we consider
the lower bound for $\omega _{o}$, i.e. $ \omega _{o}\simeq 500,$
we find that the deceleration parameter has a value close to one
half, as it should be in the Einstein-de Sitter model

Following the approach done by Uehara and Kim (1982), we may write
directly an expression for the deceleration parameter $q_{o}$
given by
\begin{equation}
q_{o}=\epsilon _{o}+\frac{\omega _{o}}{3}\epsilon _{o}^{2}+\frac{1}{2}\left(
\frac{\omega _{o}+3}{\omega _{o}+3/2}\right) \Omega _{m},
\label{q}
\end{equation}
Notice that this factor is related to the present rate of change
of the Newton's gravitational constant expressed by $\ds \left(
\stackrel{\cdot }{G}/G\right) _{o} = \mid \frac{\dot{\Phi}}{\Phi}
\mid_0 $, since $\ds G(\Phi)=\frac{1}{\Phi}$.

The contribution to the deceleration parameter in Equation
(\ref{q}) is small, since its experimental upper limit is given by
$\left( \stackrel{\cdot }{\Phi }/\Phi \right) _{o} \lesssim
10^{-10}yr^{-1}$ (Helling et al. 1983; Dickey, Newhall \& Williams
1989; Shapiro 1989). More recent, measurements on white dwarfs,
have decreased the upper limit to $\ds \mid \frac{\dot{G}}{G}
\mid_0 < 10^{-11} yr^{-1}$ (Garc\'{\i}a-Berro et al. 1995).  Still
smaller upper limits for this quantity have been reported (M\"uler
et al. 1991), where lunar laser-ranging studies of the moon's
Earth orbit yields $\ds \mid \frac{\dot{G}}{G} \mid_0 < 10^{-13}
yr^{-1}$. Since $\epsilon_0 \ll 1$, we can write, $\ds q_{o}\simeq
\frac{1}{2} \left( \frac{\omega _{o}+3}{\omega _{o}+3/2}\right)
\Omega _{m}$, which becomes exactly  one half the result of
Einstein-de Sitter's theory.


At this point we would like to mention that it is possible to
describe in the same spirit, the situation in which the model
represents an accelerating flat universe, i. e. a model in
Brans-Dicke theory in which now $\Omega_0 = \Omega_m +
\Omega_{\lambda} = 1$. But, as we will see, this case becomes
quite complicate to handle. We shall postpone the details of these
studies for the near future. We shall restrict ourselves here to
give a brief description of this situation. In this case the
condition (\ref{co1}) becomes $$
 \hspace{-3.5cm} \ds \frac{d}{d\Phi }\left(
\frac{U\left( \Phi \right) }{\Phi ^{2}} \right) +(1-3w)
\frac{U\left( \Phi \right) }{ \Phi ^{4}} $$
\be
\hspace{2.5cm}\ds =  3(1-w)\frac{\lambda}{\Phi^2 }+
3(1-3w)\frac{1}{\Phi^2 \,a^2(\Phi)}. \label{co2}
\en
If we want to get an explicit form of the Brans-Dicke scalar
potential, $U(\Phi)$, we need to know the scale factor $a$ as a
function of the Brans-Dicke scalar field $\Phi$. In order to do
this, we consider the set of basic field equations
\begin{equation}
H^{2}+H\left( \stackrel{\cdot }{\Phi }/\Phi \right) =\frac{\omega _{o}}{6}%
\left( \stackrel{\cdot }{\Phi }/\Phi \right) ^{2}+\frac{8\pi
}{3\Phi }\rho _{M}+\frac{\lambda}{3}, \label{H2}
\end{equation}
and
\begin{equation}
\stackrel{\cdot \cdot }{\Phi }+3H\stackrel{\cdot }{\Phi }=\frac{8\pi }{%
2\omega _{o}+3}\rho _{M}+\frac{2\lambda }{%
2\omega _{o}+3}\Phi, \label{Fi2}
\end{equation}
(together with equation (\ref{Q})) that can be solved exactly
(Uehara and Kim 1982). The solution for $\lambda > 0$ is given by
$$\hspace{-3.5cm}  \ds a(t) = a_o \left [ A \cosh \left( \eta
\triangle t \right ) - \frac{4 \pi}{\lambda} \right
]^{\alpha(\omega_o)} $$
\be
\label{a2} \hspace{1.0cm}\times  \left [ \frac{B \tanh
\left(\frac{1}{2} \eta \triangle t \right )- \sqrt{(4
\pi/\lambda)^2-A^2}}{B  \tanh \left(\frac{1}{2} \eta \triangle t
\right )+ \sqrt{(4 \pi/\lambda)^2-A^2}} \right
]^{\beta(\omega_o)},
\en
where,$\eta^2$, $A^2$, $B$, $\alpha(\omega_o)$ and
$\beta(\omega_0)$ are given by $\ds {\frac{2 (4 + 3 \omega_o)}{3 +
2 \omega_o }}$, $ \ds {\left [ \frac{4\pi}{\lambda}\right]
]^2-\frac{3}{2\lambda}\frac{1}{4 + 3 \omega_o}\left [
\frac{\Phi_o}{\rho_o}\right ]^2 \left [ (1+ \omega_o)\epsilon_o +
H_o \right]^2}(> 0)$,  \newline $\ds {\frac{4\pi}{\lambda} + A}$,
$\ds { \frac{1+ \omega_o }{4 + 3 \omega_o}}$ and $\ds {\frac{1}{4+
3\omega_o}\sqrt{\frac{3+2\omega_o}{3}}}$, respectively. The
interval of time,  $\triangle t = t-t_c$, is related to $\ds {t_c
= -\frac{2}{\eta \sqrt{\lambda}}\left [ \frac{4\pi/\lambda - A}{
\frac{4\pi}{\lambda} + A} \right ]}$.

Expressions (\ref{a2}) together with equation (\ref{H2}) allow us
to determine the scalar field $\Phi$ as a function of time. Thus,
in principle, we could get  the scale factor as a function of the
Brans-Dicke scalar field.

We should note that the deceleration rate $q_o$ becomes in this
case
\be
\ds q_{o}=\epsilon _{o}+\frac{\omega _{o}}{3}\epsilon
_{o}^{2}+\frac{1}{2}\left( \frac{\omega _{o}+3}{\omega
_{o}+3/2}\right) \Omega _{m} - \frac{2 \omega_o}{2 \omega_o +
3}\Omega_{\lambda}, \label{q2}
\en
with $\Omega_{\lambda}$ defined by $\ds {\frac{\lambda}{3
H^2_o}}$. Notice that if the cosmological constant contribution to
the total density parameter is significant, in agreement with the
Supernova observations, the deceleration parameter becomes
negative and thus the universe will show an acceleration. We  stop
here this analysis and, as we have mentioned above, we shall
describe in more details this situation in near future.

In going back to our simple Einstein-de Sitter-like model in BD
theory,  we  discuss in the following  some kinematical
properties.

\section{Kinematics  of the model}


\subsection{Luminosity Distance-Redshift}

The ``luminosity distance'' is defined as the ratio of the emitted energy
per unit time, ${\cal L}$, and the energy received per unit time ${\cal F}
$
\begin{equation}
d_{L}^{2}=\frac{{\cal L}}{4\pi {\cal F}}
\end{equation}
In the absence of an expansion, the luminosity distance is simply the physical
distance to the source. In closed FRM cosmology, the luminosity distance
to a source at coordinate $\left( r_{1},\theta ,\phi \right) $ is, assuming
for convenience that the observer is located at $r=0$,
\begin{equation}
d_{L}^{2}(z)=a_{0}^2 r_{1}^{2}(z)(1+z)^{2},
\end{equation}
The factor $a_{0}^2r_{1}^{2}$ in this expression is nothing but the
``inverse\ square law'', since a two-sphere surrounding the source
encompassing the observer has an area equal to $4\pi
a_{0}^2r_{1}^{2}(z)$. The factor $ (1+z)^{2}$ appears due to the
redshift of the radiation between the time of emission and the
time of detection. The parameter $r_{1}$ is determined by the expression
\begin{equation}
\int\limits_{o}^{r_{1}(z)}\frac{dr}{\sqrt{1-r^{2}}}=\frac{1}{H_{o}}%
\int\limits_{o}^{z}\frac{dz}{E(z)},
\end{equation}
where $E(z)$ is given by
\begin{equation}
\displaystyle E(z)=\frac{\left( 1+z\right) ^{\gamma \left( \omega
_{o}\right) }}{a_oH_{o}\beta \left( \omega _{o}\right) \alpha \left( \omega
_{o},H_{o}\right) }.
\label{E}
\end{equation}

Thus, we find for the Luminosity distance-redshift relation
\begin{equation}
\displaystyle d_{L}(z)=a_o(z+1)\sin \left\{ \alpha \left( \omega
_{o},H_{o}\right) \left[ 1-(1+z)^{-\beta (\omega _{o})}\right] \right\} .
\label{dL}
\end{equation}
Fig. 2 shows how the luminosity distance, $d_{L}\left( z\right) $,
changes with redshift parameter $z$, for Einstein-de Sitter model
(dotted line) and BD closed model (solid line) . From this Figure
we observe that the BD-theory begins to differ from Einstein-de
Sitter's theory at $z\simeq 100$. If we take the value for the
redshift $z$ corresponding to the value associated to the last
scattering surface, i.e. $z\equiv z_{LS}\simeq 1.100$, the
luminosity distance in Einstein-de Sitter's theory differs
approximately from its analogous expression in the BD theory by a
few percentage points.


\begin{figure}
\centerline{ \psfig{figure=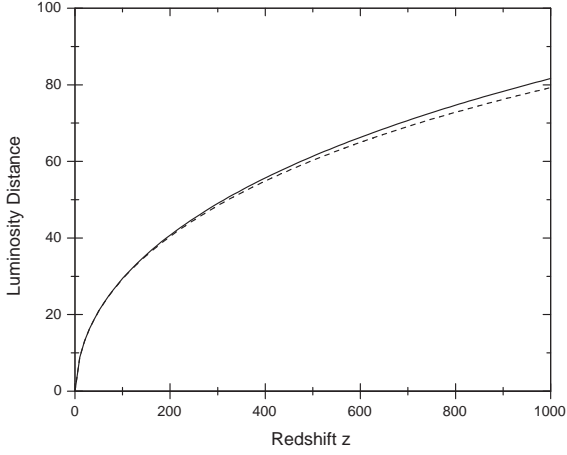,height=10.5cm} }
\caption[fig2.eps]{Plot of the luminosity distance $d_{L}(z)$ (in
unit of $ H_{o}$) as a function of the redshift $z$. The
continuous line corresponds with $\omega _{o}=500$ (BD theory) and
the dotted line represents the luminosity distance in the
Einstein-de Sitter  model. \label{fig2}}
\end{figure}


Note that for small $z$, Equation (\ref{dL}) yields
\begin{equation}
H_{o}d_{L}(z)=z+\frac{1}{2}\left( 1-\beta \left( \omega _{o}\right) \right)
z^{2}+o(z^{3}),
\end{equation}
where we have rescaled the luminosity distance by a factor of
$a_oH_o \beta \left( \omega _{o}\right) \alpha \left( \omega
_{o},H_{o}\right) .$ The first term of this expression is
nothing but the Hubble law, and from the second term, we can
read the $q_{o}$ parameter, $\displaystyle q_{o}=\beta \left(
\omega _{o}\right) =\frac{1}{2}\left( \frac{\omega
_{o}+2}{\omega _{o}+1}\right) $, which coincides with the result
obtained above.

Also, from Equation (\ref{dL}) we can use the apparent
bolometric magnitude $m(z)$ of a standard candle (with $M$ equal
to the absolute bolometric magnitude) as a function of the redshifts
\begin{equation}
\ds m(z) = M + 5 log[d_L(z)] + 25.
\label{m}
\end{equation}
If we define $D_L(z) \equiv H_o d_L(z)$, then we obtain, from
the latter Equation
\begin{equation}
\ds m(z) = {\cal M} + 5log[D_L(z)],
\label{m2}
\end{equation}
where ${\cal{M}} \equiv M-5log(H_o) + 25$, is the magnitude "zero point",
which can be determined observationally. Hamuy et al. (1996) have
determined the value ${\cal{M}} = -3.17 \pm 0.03$ using 18 supernovae
discovered by the Cal\'an/Tololo researchers. This
value is supposed to be independent of the redshift $z$.
Therefore, if we take two different values for the redshift,
we can obtain a variation of the apparent bolometric magnitude which
is given by
$\Delta m\left( z_{2},z_{1}\right) \equiv 5\log
\left[ d_{L}(z_{2})/d_{L}(z_{1})\right] $, or explicitly
$$
\hspace{-5.5cm}\Delta m\left( z_{2},z_{1}\right) =5\log
$$
\begin{equation}
\hspace{0.8cm} \ds \left\{ \frac{\left( 1+z_{2}\right)
\sin \left[ \alpha \left( \omega _{o},H_{o}\right) \left( 1-\left(
1+z_{2}\right) ^{-\beta \left( \omega _{o}\right) }\right) \right] }{\left(
1+z_{1}\right) \sin \left[ \alpha \left( \omega _{o},H_{o}\right) \left(
1-\left( 1+z_{1}\right) ^{-\beta \left( \omega _{o}\right) }\right) \right] }
\right\}.
\end{equation}
This expression could be used to restrict the value of the
BD parameter $\omega _{o}$.

By using the values $z_{1}=0.5 $ and $z_{2}=1.0$, we find the
$\Delta m$ \ increases monotonically with the BD-parameter
$\omega _{o}$. When this parameter reaches a value
close to $\omega _{o}\simeq 20$, the growth of $\Delta m$
with respect to $\omega_{o} $ increases, but slowly now,
approaching asymptotically a limiting value. At $\omega _{o} =
500$ this difference yields the value $\Delta m\simeq 3.32$.

On the other hand, if we consider the values specified by Goobar \&
Perlmutter (1995) in which the apparent magnitude (R-band) was
$m_R = 22.17 \pm 0.05$ at redshift $z=0.5$ and $m_R = 25.20 \pm
0.05$ at redshift $z=1.0$, we obtain $\Delta m_R = 3.03 \pm
0.05$. This value represents a similar order of magnitude as
that obtained for $\omega_o = 500$.


\subsection{The Angular diameter-redshift}

The angular diameter distance $d_{A}$ between a source at redshift $z_{2}$
and $z_{1}<z_{2}$ is defined by
\begin{equation}
d_{A}\left( z_{1},z_{2}\right) =\frac{a_o\sin \left[ \Delta \chi \left(
z_{1},z_{2}\right) \right] }{\left( 1+z_{2}\right) }
\end{equation}
where $\Delta \chi \left( z_{1},z_{2}\right) $ is the polar-coordinate
distance between a source at $z_{1}$ and another at $z_{2}$, in the same
line of sight
\begin{equation}  \label{D}
\Delta \chi (z_{1},z_{2})=\alpha \left( \omega _{o},H_{o}\right) \left[
\left( 1+z_{1}\right) ^{-\beta \left( \omega _{o}\right) }-\left(
1+z_{2}\right) ^{-\beta \left( \omega _{o}\right) }\right]
\end{equation}

The corresponding angular size of an object of proper length  $\ell $ at a
redshift $z$ results in $\theta \left( z\right) \simeq \ell /d_{A}\left(
0,z\right) $, which becomes (in units of $\ell H_{o}$)
\begin{equation}
\Theta \left( z\right) =\frac{1}{a_oH_{o}}\frac{\left( 1+z\right) }{\sin
\left\{ \alpha \left( \omega _{o},H_{o}\right) \left[ 1-\left( 1+z\right) %
\right] ^{-\beta \left( \omega _{o}\right) }\right\} }
\label{Theta1}
\end{equation}
Fig. 3 shows the plot of $\Theta $ as a function of $z$ in\
Einstein-de Sitter's theory and the BD theory with $\omega
_{o}=500$. These two theories coincide in the $ 0 \leq z \simeq
z_{LS}$. At $z=z_{LS}$, it is found that in the first
approximation of $1/\omega _{o}$, their difference is $\Delta
\Theta \equiv \Theta _{BD}-\Theta _{E}\sim 147\left( 1/\omega
_{o}\right) $, for any large value of $\omega _{o}$. In this plot,
we have taken $a_0 H_0 = \sqrt{2/5}$ \cite{KaTo}.


\begin{figure}
\centerline{ \psfig{figure=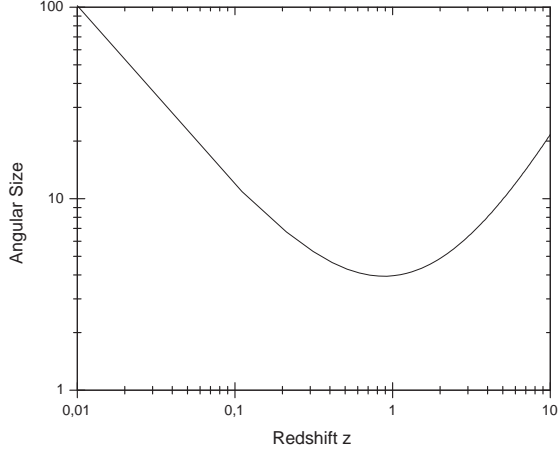,height=10.5cm} }
\caption[fig3.eps]{This plot show the angular size (in unit of $l
H_{o})$ as a function of the redshift $z$, as expressed by
Equation (\ref{Theta1}) for the case $\omega _{o}=500$. We should
mention that the corresponding curve for the Einstein-de Sitter
model is indistinguishable in the range $0 \leq z \leq z_{LS}$.
\label{fig3}}
\end{figure}


\subsection{The number count-redshift}

The number of galaxies in a comoving volume element in an angular solid area
$d\Omega $ with redshift between $z$ and $z+dz$ is sensitive to the number
of galaxies, $n$, in a comoving volume element $dV_{c}$ and the spatial
curvature:
\begin{equation}
dN_{gal}=ndV_{c}=n\frac{r^{2}}{\left( 1-r^{2}\right) ^{
{\frac12}}}drd\Omega
\end{equation}
>From this relation we can write the differential number of galaxies per
steradian per unit redshift,
\begin{equation}
\ds \frac{dN_{gal}}{dzd\Omega }\left( z\right) =\frac{n\,a_o^2\,\sin ^{2}\left[ \Delta
\chi \left( z\right) \right] }{H_{o}E(z)}
\label{n}
\end{equation}
Using the expression for $\Delta \chi (z)$ from Equation \ref{D}
and $E(z)$ from Equation (\ref{E}), we obtain $$
\hspace{-3.5cm}\ds \frac{dN_{gal}}{dzd\Omega }\left( z\right) =
n\,a_o^3 \alpha (\omega _{o},H_{o})\beta(\omega _{o})\times $$
\begin{equation}
\hspace{1.5cm}\ds \frac{\sin
^{2}\left\{ \alpha (\omega _{o},H_{o})\left[ 1-\left(
1+z\right) ^{-\beta \left( \omega _{o}\right) }\right] \right\}
}{\left( 1+z\right) ^{\gamma \left( \omega _{o}\right) }}.
\end{equation}
In Fig. 4 we have plotted the number-redshift relation (given by
Equation (\ref{n})), for $\omega _{o}=500$. Here, we have
confirmed that, for a large range of the redshift z, the function
of the number of galaxies per unit of steradian per unit redshift
in Einstein-de Sitter's model becomes almost indistinguishable
from its similar expression in the BD-theory.


\begin{figure}
\centerline{\psfig{figure=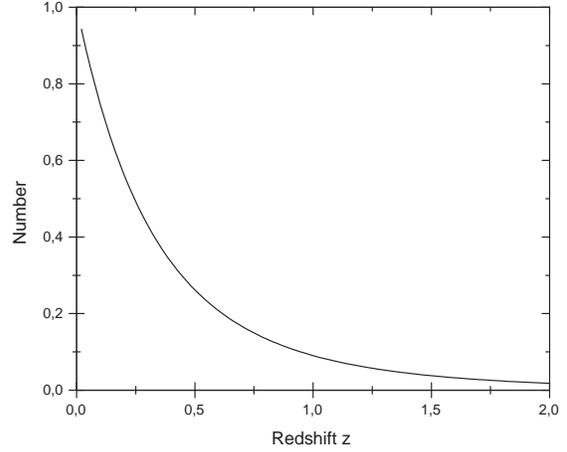,height=10.5cm}}
\caption[fig4.ps]{The differential number of galaxies per unit
redshift per steradian in units of $nH_o^{-3}$ for the BD
parameter $\omega _{o}=500$. At a low redshifts these differential
numbers rapidly approach zero. \label{fig4}}
\end{figure}


For small redshift one obtaines
\begin{equation}
\frac{dN_{gal}}{dz\,d\Omega }\left( z\right) \simeq
n\,\left( a_o\,\alpha (\omega _{o},H_{o})\,\beta(\omega _{o}) \right)^3
z^{2}+O\left(
z^{3}\right)\, .
\end{equation}
This expression indicates that, for small redshift, the slope in
the plot $\ds \frac{dN_{gal}}{dz\,d\Omega }$ v/s $z^2$ could
provide some information about the $\omega _{o}$ BD parameter,
assuming that the present value of the Hubble constant $H_o$ is
known. This, in principle, could be checked by the corresponding
observations.


\subsection{The ratio of the redshift thickness-angular size}

The redshift thickness $\delta z$ and the angular size $\delta
\theta $ of a spherical structure that grows with the expansion of the
Universe have a dependence on $z$ given by
\be
\ds \frac{1}{z}\frac{\delta z}{\delta \theta }(z)
=a_{o}H_{o}\frac{E(z)\sin \left[ \Delta \chi (z)\right] }{z},
\en
or explicitly,
\be
\frac{1}{z}\frac{\delta z}{\delta \theta }(z)
=\frac{\left( 1+z\right) ^{\gamma (\omega _{o})}\sin \left\{ \alpha \left(
\omega _{o},H_{o}\right) \left[ 1-\left( 1+z\right) ^{\beta (\omega _{o})}%
\right] \right\} }{a_oH_o\alpha \left( \omega _{o},H_{o}\right) \beta (\omega
_{o})z}
\en
Fig. 5 (lower panel) shows how
$\displaystyle\frac{1}{z}\frac{\delta z}{\delta \theta }$ changes
with redshift $z$ in the BD-theory, for $\omega _{o}=500$. We see
that, in the range of $0\leq z \leq 10$, there is no difference
when it is compared with its analogous expression in Einstein-de
Sitter model. However, at large redshift i.e. $z\equiv z_{LS}\sim
1.100$, their difference becomes significant, as we can see from
Fig. 5 (upper panel). Thus, the two expressions differ at large
redshift values.


\begin{figure}
\centerline{ \psfig{figure=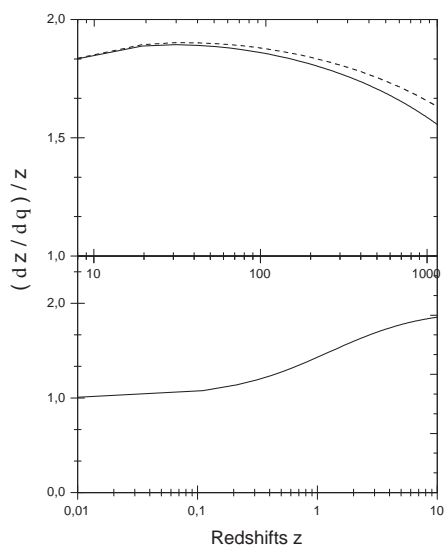,height=10.5cm} }
\caption[fig6.eps] {In the lower panel we plot $\left( \delta
z/\delta \theta \right) /z$ as a function of $z$ in BD theory (for
$\omega _{o}=500$) at low redshift. In upper panel we have plotted
$\left( \delta z/\delta \theta \right) /z$ in Einstein-de Sitter
(dashed line) and BD theories (continous line) for $z>10$. Note
that these two curves become different at very high redshift.
\label{fig6}}
\end{figure}


\section{Conclusions}

We have studied a model in which the ``quintessence'' component
was included. In this model we have chosen a Universe with closed
topology $\left( k=1\right) $ in the BD theory, and we have added
a scalar potential for the BD-scalar field. The quintessence
contribution to the matter component was fine tuned to exactly
cancel the curvature term (with the help of the BD scalar
potential) in the corresponding field Equations.

Under these conditions, different cosmological expressions were
calculated as a function of the redshift $z$. For instance, the
luminosity distant, the angular diameter, the number count and
ratio of the redshift thickness-angular size. All these quantities
were compared with their analogous expression  related to
Einstein-de Sitter model which seems flat at low redshift. In some
cases (Luminosity distance and the ratio of the redshift
thickness-angular size), the corresponding expressions become
distinguishable at a high enough redshift. At that redshift, these
differences are difficult to be directly detected . Perhaps, if we
consider other observational facts, such as, the anisotropy of the
cosmic microwave background radiation, we might say something
about these differences, and use these results for establishing a
limit for the BD parameter, $\omega_o$, since the two theories
differ in the value of this parameter.

\section*{acknowledgements}

SdC was supported from the COMISION NACIONAL DE CIENCIAS Y
TECNOLOGIA through grant FONDECYT $N^{0} $1000305 and also from
UCV-DGIP 123.744/00. NC was supported by USACH-DICYT under grant
$N^0$ 0497-31CM.


\end{document}